\documentstyle[aps,preprint]{revtex}
\tightenlines
\begin{document}

\title{Switching dynamics between metastable ordered magnetic state and
nonmagnetic ground state \\
 - A possible mechanism for photoinduced ferromagnetism -}

\author{Masamichi Nishino, Kizashi Yamaguchi}

\address{Department of Chemistry, Graduate School of Science,\\
Osaka University,       Toyonaka, Osaka 560, Japan\\} %

\author{Seiji Miyashita}

\address{Department of Earth and Space Science, Graduate School of Science,\\
Osaka University,       Toyonaka, Osaka 560, Japan}%

\maketitle
\begin{abstract}                

By studying the dynamics of the metastable magnetization of a statistical 
mechanical model we propose a switching mechanism of photoinduced magnetization.
The equilibrium and 
nonequilibrium properties of the Blume-Capel (BC) model, which is a typical model 
exhibiting metastability, are studied by mean field theory and Monte Carlo simulation.
We demonstrate reversible changes of magnetization in a sequence of
changes of system parameters, which would model the reversible photoinduced magnetization.
Implications of the calculated results are discussed in relation to the recent experimental 
results for prussian blue analogs.
\end{abstract}
\pacs{75.90.+W 64.60.My 05.70.Ce 78.20.Wc}

\section{Introduction}
During the past decade reversible changes of magnetization by magnetic field, 
electric field, and other external fields have attracted much attention not 
only because of academic interest but also because of the possibility for 
applications in devices~\cite{Hauser,Gutlich,Sci96,J.Elec97}. 
Very recently Sato, {\it et al.} have found that magnetic properties of 
the prussian blue analogs
K$_{0.2}$Co$_{1.4}$[Fe(CN)$_6$]$\cdot$6.9H$_2$O~\cite{Sci96} and
K$_{0.4}$Co$_{1.3}$[Fe(CN)$_6$]$\cdot$5H$_2$O~\cite{J.Elec97} can be
switched from paramagnetic to ferrimagnetic state (or vice versa) by visible 
and near-IR light illumination.  

According to the experiments~\cite{Sci96,J.Elec97}, Fe(II)-CN-Co(III) moieties 
in these compounds are responsible for the photo-induced effect.
In the ground state Fe and Co are in closed shell structure which is
nonmagnetic. 
However  by illumination at a wave length $\lambda_1$=500-700 nm, 
the oxidation states of Fe and Co change from Fe (${\rm t}_{\rm 2g}^6$, S=0) 
and Co (${\rm t}_{\rm 2g}^6$, S=0) to Fe (${\rm t}_{\rm 2g}^5, $ S=1/2) 
and Co (${\rm t}_{\rm 2g}^5{\rm e}_{\rm g}^2$, S=3/2), respectively. 
These magnetic moments couple antiferromagnetically, 
resulting in a ferrimagnetic magnetization at low temperatures.
The order persists after the illumination is stopped. 
Thus, the paramagnetic material is converted to a ferrimagnetic one 
by illumination and three dimensional long range magnetic ordering appears. 
After illumination at a different wave length $\lambda_2$=1319 nm,
the ferrimagnetic ordered state is switched back to the original 
nonmagnetic state.
The switching can be reproduced very reliably.
In short, these materials show two stable macroscopic states at low 
temperatures: a paramagnetic state and a ferrimagnetic ordered state and 
these bistable states can be switched reliably by using illumination. 

We consider that competition between the following two facts 
gives one of the keys to understanding this "switching" phenomenon:
The system gains energy through the effective exchange constant 
in the magnetic state which causes a long range order and 
the system loses energy through 
excitations between nonmagnetic and magnetic local moment.

This sort of macroscopic switching phenomenon is an interesting problem 
in the context of the dynamics of ordered states. 
Thus in this paper we propose a statistical 
mechanical mechanism of such switching making use of a simplified model.
In the equilibrium state, the physical 
properties must be unique functions of the system parameters. However, 
if the system has metastable states beside
the equilibrium state, the system shows a kind of hysteresis phenomenon, 
which is the key mechanism underlying the switching effect.  

The metastability is associated with a first order phase transition.
Thus we consider a model whose ground state is nonmagnetic 
and the true thermodynamic
state is paramagnetic but which has a very long-lived metastable
ferromagnetic state as an excited state.  
To this end, we here adopt the  Blume-Capel (BC) model, 
which is a ferromagnetic $S=1$ Ising model ($S^z=\pm 1$ or 0) with 
a crystal field splitting for easy-planar symmetry:
\begin{equation}
{\cal H}= -J \sum_{\langle i, j \rangle} S_iS_j + D \sum_{i}{S_i}^2,
\end{equation}
where $J$ is the exchange constant and $D$ is the crystal field splitting 
and $ {\langle i, j \rangle}$
indicates that the summation is over all nearest-neighbor pairs on a lattice. 
We study the model on the simple cubic lattice with the linear dimension $L$.
This $D$ express the excitation energy from a nonmagnetic state to a
magnetic state.
Modeling the above mentioned mechanism, the spin $S_i$ at a site represents 
a local  magnetic state ($S_i=\pm 1$) or nonmagnetic state ($S_i=0$).  
We deal with ferromagnetic interaction ($J>$ 0) and positive $D$.
The BC model was originally proposed as a way to study first-order 
magnetic phase transitions~\cite{BC1,BC2}. 

In this model the coupling between $S_i$s is ferromagnetic while 
the magnetic moments are coupled antiferromagnetically 
in the above mentioned material. However, the inherent mechanism of the 
switching originates in the existence of a metastable ordered state
due to the competition between magnetic interaction ($J$) and local
excitation energy ($D$).
Thus, the modeling by the BC model does not lose the essential physics 
and we choose this model as the simplest one which describes the essential 
mechanism for the reversible magnetic switching. 
More complicated models for individual systems could be provided if necessary. 

In order to switch between the bistable states of the model, we make 
use of a change of physical properties due to the illumination.
When photo-sensitive materials are illuminated, we generally expect that 
their physical properties change.  The Kerr effect is a typical example, where 
the refractive index changes with the amplitude of the light.  
We expect that the parameters of the system, such as $D$ and the temperature 
$T$ would be changed during the illumination: $(D, T)$ $\rightarrow$ $(D', T')$.
The amount of the change would depend on the frequency of the light and/or 
on the amplitude of the light, {\it etc}..
The state of the system changes to that for $(D', T')$.
When the illumination is stopped the parameters will be back to the original 
values: $(D', T')$ $\rightarrow$ $(D, T)$. Contrary, the state may not come 
back to the original state because of the bistability. 
Thus such illumination would provide a switching procedure.

Using a Monte Carlo simulation, we demonstrate switchings of the 
magnetization in a sequence of changes of parameters $(D, T)$, 
which corresponds to the
reversible switching.  We predict various features of the switching, 
such as, a relationship between the magnetization during the illumination and 
the state after the illumination.  
According to the effective parameters of the system during the illumination, 
the state after illumination is determined to be magnetic or nonmagnetic. 
However, if the effective parameters take intermediate values, 
the state after the illumination becomes very stochastic. 
Namely, the switching to an desired state may fail with some probability.
This uncertainty of the switching is also studied.

In Sect. II, we study the thermodynamic properties of the BC model and
the nature of the metastability of the model. 
In Sect. III, a mechanism of the reversible switching is proposed.
Summary and discussion are given in Sect. IV.

\section{Thermodynamic properties of the BC model}

In this section we study the thermodynamic properties of the BC model and
the nature of the metastability of the model.

\subsection{Phase transitions and phase diagram}
Thermodynamic properties of the BC model have been studied extensively, 
{\it i.e.}
the model including the extended BC model have been studied by the mean-field 
approximation~\cite{BC1,BC2,BC3},
renormalization~\cite{Mahan,P.R.B86,P.R.B97},
and Monte Carlo methods~\cite{Tanaka,Tanaka2}.
In particular, the tricritical point of the model between the second order
transition line and the first order transition line has been investigated
by various methods~\cite{Stauffer,Jain,Landau,P.R.E97}.
In Fig. 1 we depict the phase diagram in the $(D,T)$ plane, 
where the points indicated by circles
have been determined by Monte Carlo simulations.
The closed circles are determined from the cross point in the Binder plot
of the magnetization~\cite{Binder} and we find the phase transitions are of the
second order. 
The open circle shows a first-order phase transition point.
Although there are several methods to determine the first
order phase transition point making use of histogram of the order
parameter~\cite{Kosterlitz}, we use the following 
simplified method.
First we look for the hysteresis:  We perform a simulation with
a disordered initial state and decrease the temperature gradually,  and find a
temperature, (say, $T_{\rm L}$),
at which the system jumps to the ordered state. Then we perform another
simulation with a complete ordered initial state at a low temperature 
and increase the temperature gradually to find a temperature ($T_{\rm H}$)
at which the system jumps to the disordered phase.
If $T_{\rm L}$ and $T_{\rm H}$ are separated significantly, 
then we regard the phase transition as first order.
In order to determine the critical temperature 
we performed the following simulations:
We prepare an initial configuration in which half of the system is in the
ordered state and the other half is in the disordered state. 
We perform Monte Carlo simulations
with different sequences of random numbers (in most cases we used 30 samples).  
If all the samples reach to the ordered phase, then we regard the set of
the parameters $(D,T)$ as belonging to the ordered state.
 On the other hand if all the samples reach the disordered phase, then it
is regarded as the disordered state.
If the samples distribute among both phases, we suppose we are in the
critical region.
The error bar in Fig. 1 shows the maximum range of this critical region and the 
open circle is put in the middle of this region.
This procedure has been done for $L=10$. If we increase the size of the
system the
error bar would shrink. However for the purpose of the present paper,
we need only a rough phase diagram and Fig. 1 is satisfactory in this sense.

The triangle shows the position of the tricritical point obtained by M.
Deserno~\cite{P.R.E97}, and our results are consistent with that in Fig. 1. 
This in turn demonstrates the reliability of present computations.

\subsection{Metastability of the BC model }

As has been mentioned in the introduction, the BC model has metastable state.
First we investigate the free energy obtained by a mean-field approximation
as a function of the magnetization:
\begin{equation}
F(M)=-k_{\rm B}T\log[{\rm Tr}\exp(-\beta{\cal H}_{\rm MF}(M))]+{1\over2}
{\beta}JzM^2
\end{equation}
with
\begin{equation}
{\cal H}_{\rm MF}(M)=zJMS-DS^2,
\end{equation}
where ${\beta}$ is $1/ k_{\rm B }T$, $z$ is coordination number and $M$ is the mean
magnetization per site. For the simple cubic lattice, $z=$ 6. 
The free energy is explicitly given by\
\begin{equation}
F(M) = -k_{\rm B}T~{ \log[ 2 \exp( -{\beta}D)\cosh ({\beta}JzM) +1] 
+{1\over2} {\beta}JzM^2 }.
\label{Fmf}
\end{equation}

Hereafter we take $J$ as a unit of energy and also we put
$ k_{\rm B}= 1$.
For $D=$ 1.0 and $T=4.0$
in the paramagnetic region, $F(M)$ has a single minimum
(Fig. 2(a) for $D=$ 1.0 and $T=4.0$).
If $D$ is small the system shows the second-order phase transition
and $F(M)$ has double minima (Fig. 2(b) for $D=$ 1.0 and $T=1.2$). 
On the other hand if $D$ is close to 3, then the system
shows a first order phase transition and $F(M)$ has 3 minima;
for $D>3$ the nonmagnetic state is the true equilibrium state 
(Fig. 2(c) for $D=$ 3.2 and $T=$1.0), and
for $D<3$ the ferromagnetic state is the true equilibrium state 
(Fig. 2(d) for $D=$ 2.8 and $T=$1.2).
If $D$ is large the system is nonmagnetic for all temperature 
and $F(M)$ has a single minimum.

In Fig. 3, we show a phase diagram including the metastable region in the
mean field free energy (\ref{Fmf}). 
For $D$ which is smaller than that at the tricritical point ($D_{\rm
t},T_{\rm t}$), 
the shape of the free energy changes as Fig. 2(a) (the paramagnetic region,
hereafter we refer to this region as I.)
 $\rightarrow$ Fig. 2(b) (an ordered state, region II ) $\rightarrow$ 
Fig. 2(d) (an ordered state with a metastable paramagnetic state, region III).
The tricritical point ($D_{\rm t},T_{\rm t}$) is found to be (2.7726,
2.0)~\cite{BC3} and plotted by a triangle in Fig. 3.
The boundary between I and II is the second order equilibrium  
phase transition line.
This boundary is shown by a solid line. On the other hand the boundary between 
II and III does not correspond to any equilibrium phase transition
but it shows a point where the metastability appears.
This boundary is shown by a dash-dotted line.
On these lines the coefficient of $M^2$ of $F(M)$ vanishes.
These lines join at a point, say Q, $(D_{\rm Q},T_{\rm Q})=(2.7783, 1.899)$. 
The value of $D$ at this point is larger than $D_{\rm t}$.
For $3>D>D_{\rm Q}$, the shape of the free energy changes as
Fig. 2(a) $\rightarrow$ Fig. 2(c) (a paramagnetic state with a metastable
ferromagnetic state, region IV) $\rightarrow$ Fig. 2(d).  
The boundary between I and IV is where the metastability appears,
which is shown by a dotted line.
The boundary between III and IV is the first order phase transition line 
in the equilibrium, which is shown by a dashed line.
As shown in the inset in Fig. 3,
between $D_{\rm t}$ and $D_{\rm Q}$, complicated changes of the shapes of
$F(M)$ occur:
(a) $\rightarrow$ (c) $\rightarrow$ (d) $\rightarrow$ (b) $\rightarrow$ (d).
For $D>3$, only the boundary (dotted line) for the metastability exists. 

Second, we investigate the phase diagram for the metastability 
corresponding to Fig. 3 by a Monte Carlo simulation.
We performed simulations with a complete ferromagnetic initial configuration
and counted how many samples decay to the disordered phase within 
10,000, 100,000 and 1,000,000  Monte Carlo Steps (MCS) in the system of
$L=10$ and 20
with the periodic boundary condition.

In Fig. 4, we plot the highest temperature at which more than $2/3$ of 
samples ($L$=10 and 30 samples) remain in a ferromagnetic state with 
$|\sum_iS_i|>
0.5L^3$ after 10,000 MCS by upward triangles,  after 100,000 MCS by circles and 
after 1,000,000 MCS by downward triangles. 
Here we find that the transition region between simple paramagnetic region and
metastable region is rather narrow and we can distinguish the region of
metastable
state rather clearly.
The data for 1,000,000 MCS shows that the transition is not sharp near $D=4.0J$.
In the inset of Fig. 4, the number of samples which remain
ferromagnetic are shown.

Here, we investigate the boundary of the metastability of the complete
ferromagnetic state from a view point of local nucleation process.
Let us consider configurations with a cluster of nonmagnetic sites. 
The energy difference between the complete ferromagnetic state and a state
with a cluster, say $\Delta E$, is given by 
\begin{equation}
\Delta E = -n D + m J
\end{equation}
where $n$ is the number of nonmagnetic sites and $m$ is the number of
excited bonds ({\it i.e.}, $S_iS_j=0$).
In Fig. 5(a) the excess energy, $\Delta E$, of configurations with a
cluster of nonmagnetic sites are shown.
We found easily that the complete ferromagnetic state is unstable even for
a flip of a single nonmagnetic site when 6$J<D$.
Thus the upper limit of the metastability locates at $D=6J$.
For $3J>D$, the ferromagnetic state becomes the true equilibrium state.  
Thus the metastable ferromagnetic state exists in the range $3J<D<6J$, 
which is consistent with the above investigations.
At a given value of $D$ in this range, the excess energy of a cluster
configuration, $\Delta E$, increases as $n$ increases for small values of
$n$. 
The complete ferromagnetic state is
locally stable against fluctuation with such clusters . 
On the other hand for large values
of $n$, $\Delta E$ decreases when $n$ increases. 
Between these two regions, $\Delta E$ has a maximum. 
The configuration for the maximum $\Delta E$ is called critical nucleus and
$n$ at this configuration is named $n_{\rm C}$. 
This means that once a cluster of nonmagnetic sites larger than 
$n_{\rm C}$ appears it grows and
the ferromagnetic state is destroyed. 
In Fig. 5(b) the minimum $\Delta E$ for each $n$ is plotted for various values of $D$. 

The boundary of the metastability may be given by
\begin{equation}
e^{-\beta \Delta E(n_{\rm C})} \simeq p_{\rm min},
\end{equation}
where $p_{\rm min}$ is the smallest nucleation rate which is detectable in
the observation (in Monte Carlo simulation in the present case).
Because we are studying phenomena in the time scale
$t<10^6$  MCS and the size of system $L^3\simeq10^3$, 
we could find phenomena of probability $p_{\rm min}\simeq 10^{-9}$.
Because we are interested in  the temperature region, $T<0.5$,
only the phenomena with $\Delta E$ of
\begin{equation}
\Delta E < 10J
\end{equation}
are meaningful in the present observation because 
\begin{equation}
e^{-\beta \Delta E} > p_{\rm min} \simeq 10^{-9}.
\end{equation}

For $D>4J$, the ferromagnetic state is only stable against clusters with a
few nonmagnetic sites. Thus the region of metastability exists only if $T \ll
1$.
In the range ($\Delta E < 10J$) we find that the $\Delta E(n)$ becomes very
flat at $D \simeq 4$.
Thus we expect that the boundary of the metastability becomes ill defined
for this case, which would give the explanation of the wide range of the
transition observed in Fig. 4.

In the same way, we determine the metastable region for the disordered
phase which is also shown in Fig. 4 ($D<3$),
by  boxes (10,000MCS) and diamonds(100,000MCS). Qualitatively,
the phase diagram in Fig. 4 agrees well with that of Fig. 3.
If we look at Fig. 4 carefully, the boundary corresponding to the second
order phase transition line in Fig. 1 (dotted line) shifts to low
temperature side.
This disagreement is simply due to the definition of the 
boundary in Fig. 4 as discussed in Appendix. 

Next we study the size dependence of the metastability of the 
ferromagnetic state.
We expect the metastable state in the present model is in the so-called 
stochastic region or single droplet region~\cite{Rikvolt}.
There 
we expect the nucleation rate in the whole volume is about 8 times larger
in the system of $L=20$,
because the volume of the system of $L=20$ is 8 times larger than that for
$L=10$. 
Actually in the cases with $D>$ 3.0, we  observe the marginal points
for 12,500MCS (the symbol $\times$ in Fig. 4) and 125,000MCS (the symbol +
in Fig. 4) of the systems of  $L=20$
well overlap with those for 100,000 and 1,000,000 in the system of $ L=10$.

\section{Reversible switching}

In this section we consider a possible mechanism of switching between 
the ordered state and disordered state, making use of the structure of
metastability studied in the previous sections.
We consider a system which is nonmagnetic in the equilibrium state 
but which has a long-lived metastable ordered state.
We can find such a system in the BC model with $D$ a little bit larger than 3.0
at low temperature. 
We take a point A ($(D_{\rm A},T_{\rm A})=(3.2,0.6)$) as such a point 
(see Fig. 6).
Here simulations are done in a system of $L=10$.
We have checked that qualitative features do not change in the system of
$L=20$ and
even quantitatively most properties are reproduced in the system of $L=20$.
If we start with a high temperature ($T=1.5$) and cool down the system, we find
a paramagnetic state at A. On the other hand if we start with complete
ferromagnetic
state at $T=0$ and warm up the system we find a ferromagnetic state at A.
If the system is put in other environments, such as under illumination,
the parameters, $J$ and $D$, would be
renormalized.
Let a point X $(D_{\rm X},T_{\rm X})$ be such a state with renormalized 
parameters.
Now we study the change of magnetization in the change of the parameters
A $\rightarrow$ X $\rightarrow$ A. Here let us take $D_{\rm X}$ to be 2.8.
We consider the temperature $T_{\rm X}$ in the ordered state. 
Here we assume $T_{\rm X}$ is above the temperature where the
metastable paramagnetic state exists. 
When the system move to X, the system rapidly becomes ferromagnetic.
If $T_{\rm X}$ is low enough, only clusters of nonmagnetic sites 
which are smaller than the critical nucleus for $D_{\rm A}$ exist in X.
After the system comes back from X to A, the system is trapped in the 
metastable ferrimagnetic state. 
As an example of such $T_{\rm X}$ we take $T_{\rm X}=1.25$ and
the change of magnetization is shown in Fig. 7. Hereafter we call this point B;
$(D_{\rm B},T_{\rm B})=(2.8,1.25)$.
There we repeat the process 
(A $\rightarrow$ B $\rightarrow$ A $\rightarrow$ B $\cdots$) 
in order to check the stability of the dynamics.

Here simulations were performed as follows:
First we simulate the system at a high temperature $T=1.5$ with 50,000MCS 
and then gradually reduce the temperature by $\Delta T=0.1$ iteratively 
until $T$=0.6. 
At each temperature 50,000 MCS are performed.
Next the system moved to the point B and 20,000MCS is performed. 
Then the system comes back to A and there another 100,000MCS is performed, 
where we find that the system is always in the metastable ferromagnetic state.

Next, we take  $T_{\rm X}$ at a higher temperature $T_{\rm X}$=1.5, 
where many clusters of nonmagnetic sites are exited.
This point will be called C; $(D_{\rm C},T_{\rm C})=(2.8,1.5)$.
The change of magnetization is shown in Fig. 8, where the system
at point A is always paramagnetic.

Whether the state after the system comes back from
X is ferromagnetic or paramagnetic 
depends on the temperature $T_{\rm X}$. 
For intermediate temperatures, the state after coming back from X 
distributes among ferromagnetic or paramagnetic state.
For example the time evolution of the magnetization for $T_{\rm X}=1.38$ is
shown in Fig. 9. 
We investigate the reliability of the switching by making use of the quantity:
\begin{equation}
P = \left({ N_{\rm f}-N_{\rm p} \over N_{\rm f}+N_{\rm p} }\right)^2,
\end{equation}
where $N_{\rm f}$ is the number of appearance of ferromagnetic state
and $N_{\rm p}$ is as well for paramagnetic state. 
This quantity indicates the degree of certainty of the state after the
system comes back from X. 
If there exists in X a cluster of nonmagnetic sites which is larger 
than the critical size for $D_{\rm A}$, the state will become nonmagnetic after coming back to A.

The distribution of the size of nonmagnetic cluster in X, $p_{\rm X}(n)$,
determines the distribution of $N_{\rm f}$. 
The quantitative analysis for $p_{\rm X}(n)$ is difficult but we expect
that $N_{\rm p}/(N_{\rm f}+N_{\rm p}$) is a monotonic function of $T_{\rm
X}$ and is very small for small $T_{\rm X}$ and $N_{\rm f}/(N_{\rm
f}+N_{\rm p}$) is also very small for large $T_{\rm X}$. 
Thus we identify the three regions of $T_{\rm X}$, {\it i.e.}, $N_{\rm
p}/(N_{\rm f}+N_{\rm p}) \simeq 0~ (P=1)$, 0 $<$ $N_{\rm p}/(N_{\rm
f}+N_{\rm p})$ $<$ 1 (0 $<$ $P$ $<$1) and $N_{\rm p}/(N_{\rm f}+N_{\rm p}) 
\simeq 1~(P$=1) rather clearly.
In the figure we also plot the square of magnetization per site, 
$\langle M^2\rangle/L^3$, during the illumination.
The $T_{\rm X}$-dependence of the magnetization is rather mild while the
change of $N_p$ is sharp.

We estimate $P$ as a function of $T_{\rm X}$, counting $N_{\rm f}$ 
in continuous 10 times repetition: 
A $\rightarrow$ X $\rightarrow$ A $\rightarrow$ X $\cdots$.
We performed simulations for each $T_{\rm X}$ value with
different random number sequences. 
The average of $P$ is shown in Fig. 10 estimated from four sets of
$N(=N_{\rm f}+N_{\rm p}$)=10 samples varying $T_{\rm X}$ from 1.2 to 1.5.
In Fig. 10, $P$ is almost 0 near $T$=1.35 and this temperature indicates the
marginal point for the switching and there large fluctuations of $N_{\rm f}$ 
occur, while in the range $T\le1.25$ or $T\ge1.45$, 
the switching is very reliable, i.e. $P\simeq$ 1. 
 
Let us now demonstrate the reversible switching.
In order to realize reliable switchings we choose the points, B and C.
The dynamics of magnetization is shown in Fig. 11 for A $\rightarrow$ B 
$\rightarrow$ A $\rightarrow$ C $\rightarrow$ A $\rightarrow$ B 
$\rightarrow$ A $\rightarrow$ C $\cdots$, where the state at the point A 
is ferromagnetic (F) after coming back from the point B and it is 
paramagnetic (P) after coming back from the point C.
We regard the point B as the state during illumination with the frequency 
$\nu_{1}$ and the point C as the state during illumination with the frequency 
$\nu_{2}$. The state at the point A is switched by the illuminations 
of the frequencies $\nu_{1}$ and $\nu_{2}$ from P to F and from F to
P, respectively. 
In this way the reversible magnetization process is demonstrated.
If the temperature at the point B becomes lower and the temperature at C
becomes higher, the switching becomes more reliable.

As reported in the experiment~\cite{Sci96,J.Elec97},  light with shorter wave
length was used for switching to the ferromagnetic state.
Generally we expect that illumination at short wave length causes large
renormalization of the parameter, although the renormalization depends 
on the microscopic properties of the individual material.
In the above demonstration, we chose a higher temperature to switch off the 
magnetization. 
Intuitively this choice is not consistent with the experimental situation.
Thus we demonstrate the switching taking another point for C.
Here we take the point $(D=3.1,T=1.0)$ instead of C.  
This new point will be called C'. 
It would be plausible that the renormalized values of
$D$ and $T$ shift from A to B through C' with a change of the frequency of the light.
The dynamics of magnetization is shown in Fig. 12 for A $\rightarrow$ B $\rightarrow$ 
A $\rightarrow$ C' $\rightarrow$ A $\rightarrow$ B $\rightarrow$ A $\rightarrow$ C' $\cdots$, 
where we also find steady reversible switching.

\section{Summary and Discussion}

We have shown that switching of macroscopic states is possible in a sequence of 
parameter changes of the system, which would model 
reversible switching 
in the experiments of photoinduced ferrimagnetism.
So far the switching mechanism has been discussed with
the picture of the adiabatic potential for the nonmagnetic ground state
and magnetic excited state in a microscopic structure.
Such microscopic structure is the necessary condition for the reversible 
switching. 
In order to explain the change of macroscopic state, 
we need to understand how the macroscopic order parameter behaves.

We proposed that the existence of metastability due to the
competition between the magnetic coupling and local excitation energy
gives an essential mechanism of such a switching.
We have investigated thermodynamic properties of the BC model. In 
particular the metastable ferromagnetic region of the model was studied
in detail by a mean field theory and a Monte Carlo method. 
The metastable ferromagnetic region was found to be very long-lived.
We took this model to explain the reversible switching phenomena.
On the assumption that photon's effect causes renormailzations of $(D,T)$,
we observed the dynamics of magnetization for variety of changes 
of the parameters $(D,T)$.  
We demonstrated that in suitable changes of $(D,T)$
the state of the system can be switched to be a different one even if the
parameters of the system come back to the original values, which is the
essential property of the reversible magnetic switching.
The present study thus provides a statistical mechanical mechanism 
for such switching. 
We found thermodynamical properties of the switching,
such as dependence on the temperature and field, {\it etc}.
and the reliability of the switching.

In this paper we studied very simplified model but we could provide more
close models to experimental situations. For example,
${\cal H}= -J\sum_{\langle ij\rangle}S_j\sigma_i\tau_j\mu_i
+D_{\tau}\sum_{i\in {\rm A}} \tau_i^2 
+D_{\mu}\sum_{i\in {\rm B}}\mu_i^2$ $(J<0$(antiferromagnetic), $D_{\tau}>0, 
D_{\mu}>0 )$ would be more closely describe the prussian blue analogs, 
where $S_i=\pm 3/2,\pm 1/2$ and $\sigma_i=\pm 1/2$. 
$\tau$ and $\mu$ take 0 or 1, representing the nonmagnetic or
magnetic state. Here A and B denote the sublattices of the lattice.
The $\langle ij\rangle$ denotes all nearest neighbor pairs.
Such specification
will be investigated with detailed experimental data for the material 
in the future.
Furthermore making use of other models with the first order phase transition, 
we could provide models for different kinds of switchings, 
such as a switching between several values of magnetization, ${\it etc.}$.

In order to study the phenomena microscopically, we have to know how the 
parameters $(D, T)$ are renormalized during the illumination. 
Dependence of the parameters on the frequency and/or the amplitude of 
light is challenging problem which we would like to study in the future.

\acknowledgements

The authors would like to thank Dr. O. Sato for his kind discussion
on the experiments and also thank Professor K. M. Slevin for his kind critical 
reading of the manuscript.        
The present work was supported by Grant-in-Aid for Science
Research from the Ministry of Education, Science and Culture of Japan.

\appendix

\section{The spontaneous magnetization}

We determined the boundary in Fig. 4 as the highest
temperatures
at which the system does not have the magnetization below $M=0.5L^3$ 
during the simulation.
Thus if the equilibrium spontaneous magnetization $M_{\rm s}(T)$ is less than 
$M=0.5L^3$, the temperature $T$
belongs to the right hand side region of the boundary ($T$ is higher than the 
boundary), although this temperature $T$ is lower than
the phase transition point $T_{\rm C}(D)$. 
Furthermore,
even if the equilibrium magnetization is larger  than $M=0.5L^3$, the system
can have the magnetization of $M<0.5L^3$ as a fluctuation.
Thus the boundary in the present criterion locates in the low temperature
side of the true phase boundary.

Generally the spontaneous magnetization $M_s$ changes very
rapidly with
the temperature and the point at which $M_{\rm s}=0.5L^3$ is very close 
to the phase transition point. For the two dimensional Ising model on the square lattice, 
the spontaneous magnetization per
spin is given by C. N. Yang's well-known solution~\cite{Yang}:
$m_{\rm s}=(1-({\rm sinh}2{\beta}J)^{-4} )^{1/8}$.
Here $k_{\rm B}T_{\rm C}/J$=2.2692 and $k_{\rm
B}T_{1/2}/J$= 2.2674 for $m_{\rm s}$=0.5.
This temperature for $m_{\rm s}$=0.5 is very close to $T_{\rm C}$:
${T_{1/2}}/{T_{\rm C}}$=0.999.
The present model is three dimensional and the change of magnetization is milder than 
that of 2D because the exponent $\beta$ is about 0.32 instead of 1/8. 
However, $m_{\rm s}(T)$ shows a very sharp change around $T_{\rm C}$ as shown 
in Ref.~\cite{3DIsing}.

Thus the difference discussed above is mainly due to the latter reason, 
{\it i.e.} a fluctuation of the magnetization.
Actually the boundary in the system of $L=20$ it is found to 
locate more close to the phase transition point.

\newpage

\begin{figure}
\caption{ The phase diagram of the BC model.
The closed circles denote the second ordered phase transition obtained by
a Monte Carlo simulation. The triangle denotes the tricritical point
determined by M. Deserno.
The open circle denotes the first order phase transition obtained by
a Monte Carlo simulation. The error bar denotes the region of the hysteresis. }
\end{figure}

\begin{figure}
\caption{The mean field free energies of the BC model for various parameter
sets $(D,T)$.} 
\end{figure}

\begin{figure}
\caption{The phase diagram including the metastable region in the mean field
theory (See text). }
\end{figure}

\begin{figure}
\caption{The metastable ferromagnetic and paramagnetic region obtained by
Monte Carlo
method. The dashed line is the phase boundary shown in Fig. 1. The symbols
denote
the boundaries of the metastable region (For details, see text). }
\end{figure}

\begin{figure}
\caption{ (a) Clusters of $n$ nonmagnetic sites and the energy difference
($\Delta E$) from the completely ferromagnetic state. (b) Dependence of 
($\Delta E$) on the number of sites $n$ for various values of $D$. 
Symbols +, $\bigtriangledown$, $\times$, $\Box$, $\bullet$,
$\bigtriangleup$, $\Diamond$, and $\circ$ denote data for D=1, 2, 3, 3.5,
4, 4.5, 5, and 6 respectively.
The maximum point corresponds to the size of the critical nucleus for each
value of $D$. } 
\end{figure}

\begin{figure}
\caption{Switching path in $(D,T)$ plane with phase boundary and metastable
regions. Definitions of A, B, C and C' are given in the text.}
\end{figure}

\begin{figure}
\caption{The change for the square of magnetization per site
$\langle M^2\rangle/L^3$ 
in the case of $T_{\rm X}$=1.25.}
\end{figure}

\begin{figure}
\caption{The change for the square of magnetization per site
$\langle M^2\rangle/L^3$ 
in the case of $T_{\rm X}$=1.5.}
\end{figure}

\begin{figure}
\caption{The change for the square of magnetization per site
$\langle M^2\rangle/L^3$ 
in the case of $T_{\rm X}$=1.38.}
\end{figure}

\begin{figure}
\caption{The closed circles denote 
average of the reliability $P$ as a function of
$T_{\rm X}$ for fixed $D_{\rm X}$=2.8. The triangles denote
the square of magnetization per site
$\langle M^2\rangle/L^3$.}
\end{figure}

\begin{figure}
\caption{The change for the square of magnetization per site
$\langle M^2\rangle/L^3$ 
in the case of
choosing B=$(D=2.8,T=1.25)$ and C=$(D=2.8,T=1.5)$.}
\end{figure}

\begin{figure}
\caption{The change for the square of magnetization per site
$\langle M^2\rangle/L^3$ 
in the case of
choosing B=$(D=2.8,T=1.25)$ and C'=$(D=3.1,T=1.0)$.
}
\end{figure}

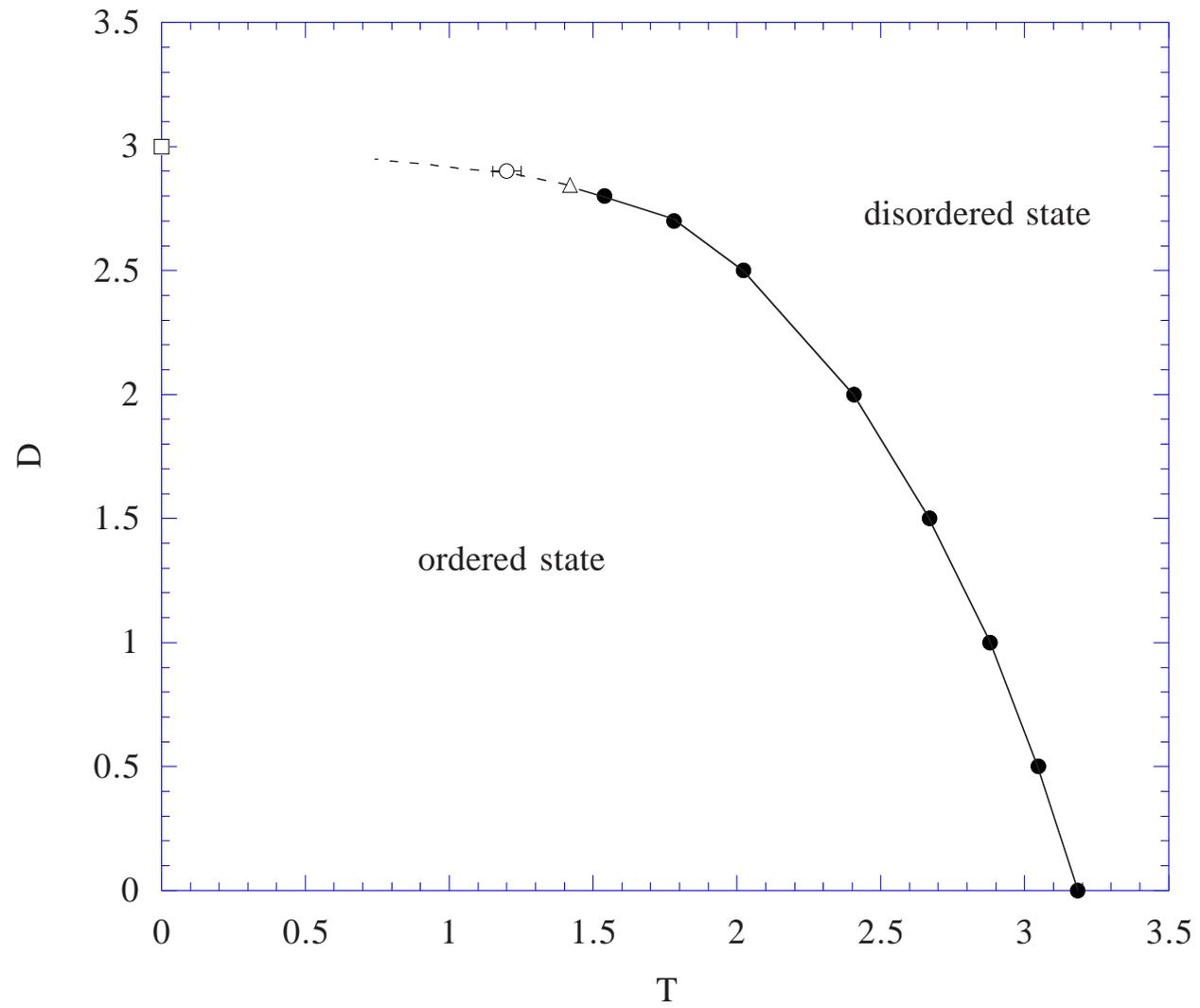
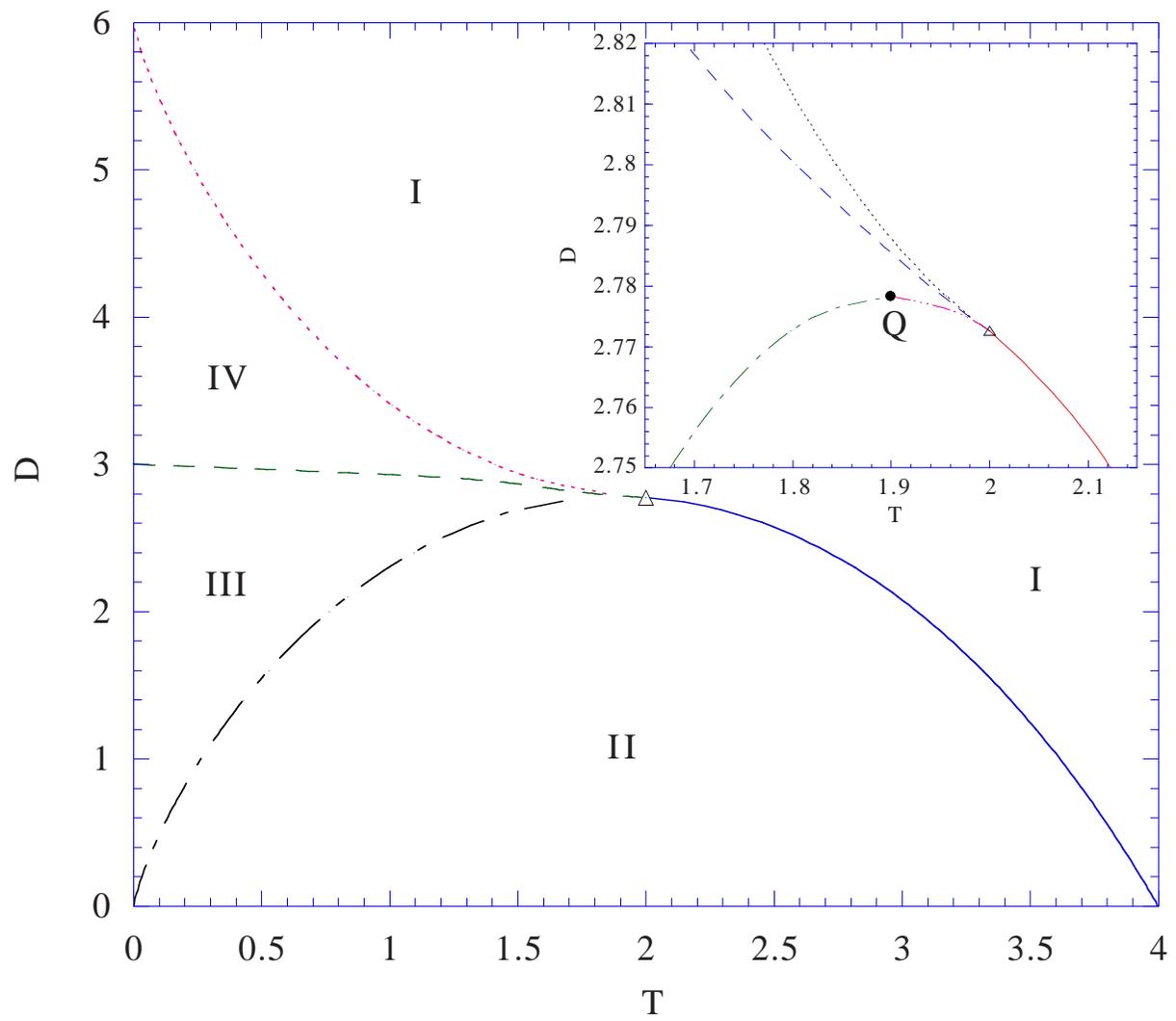
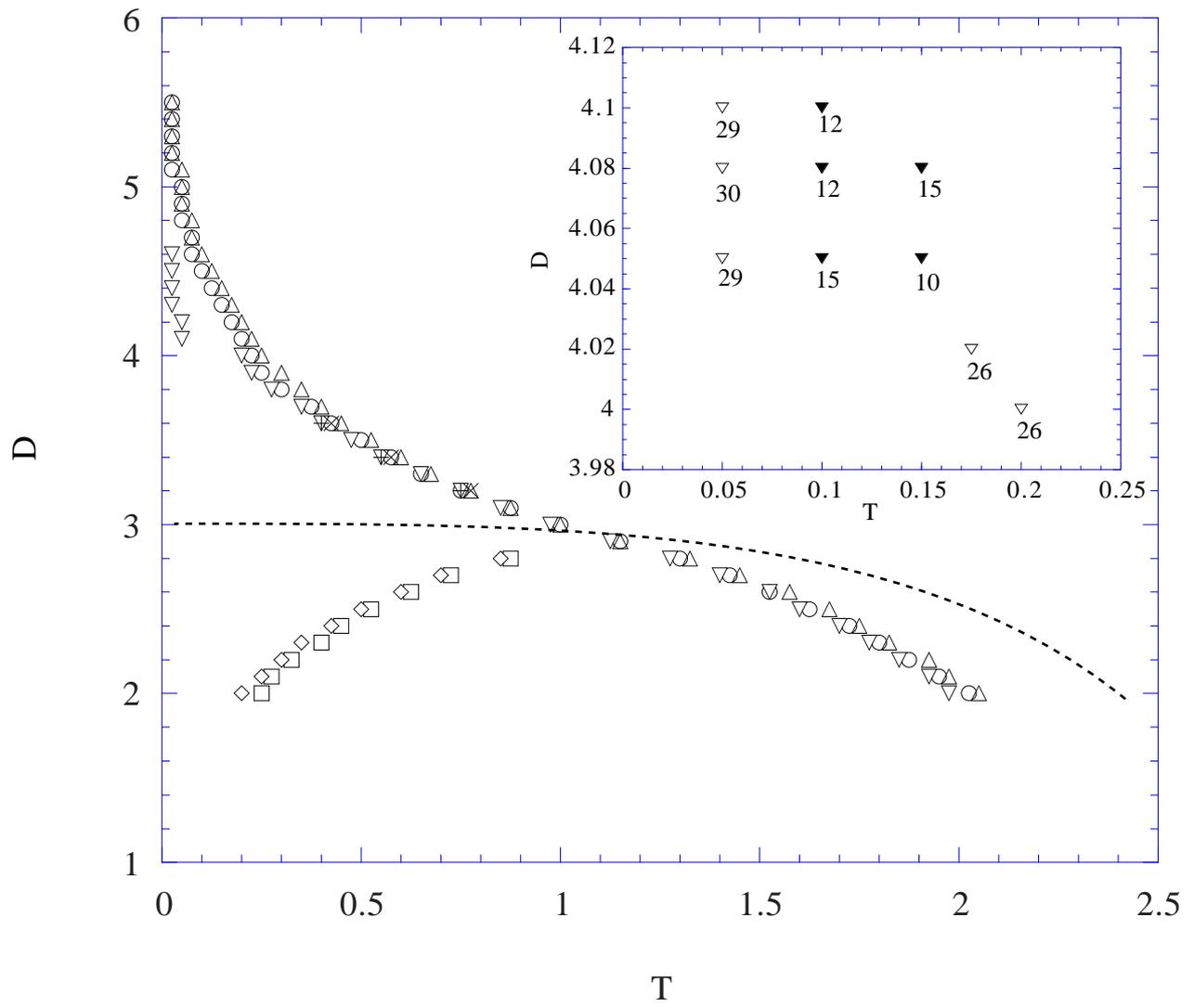
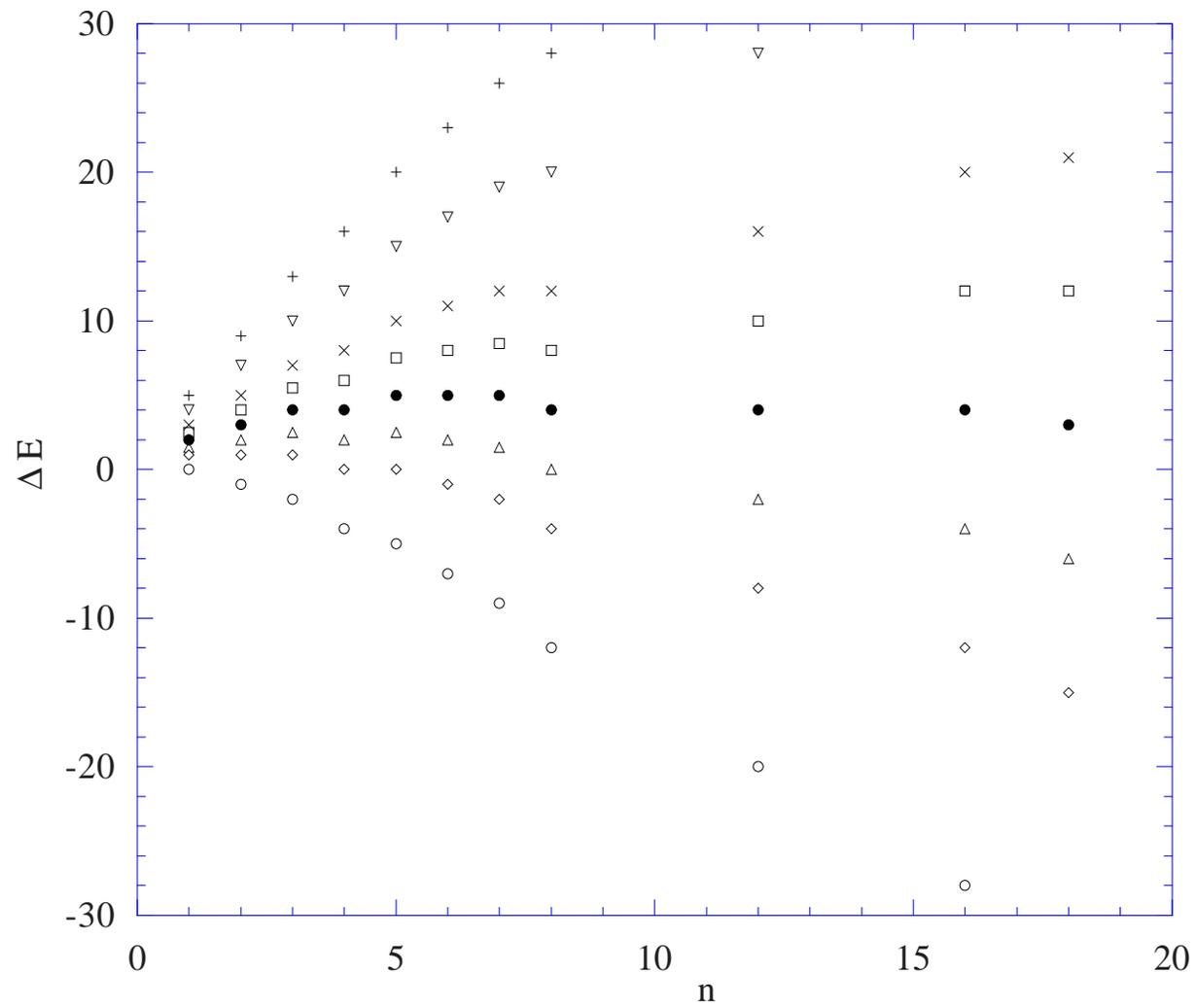
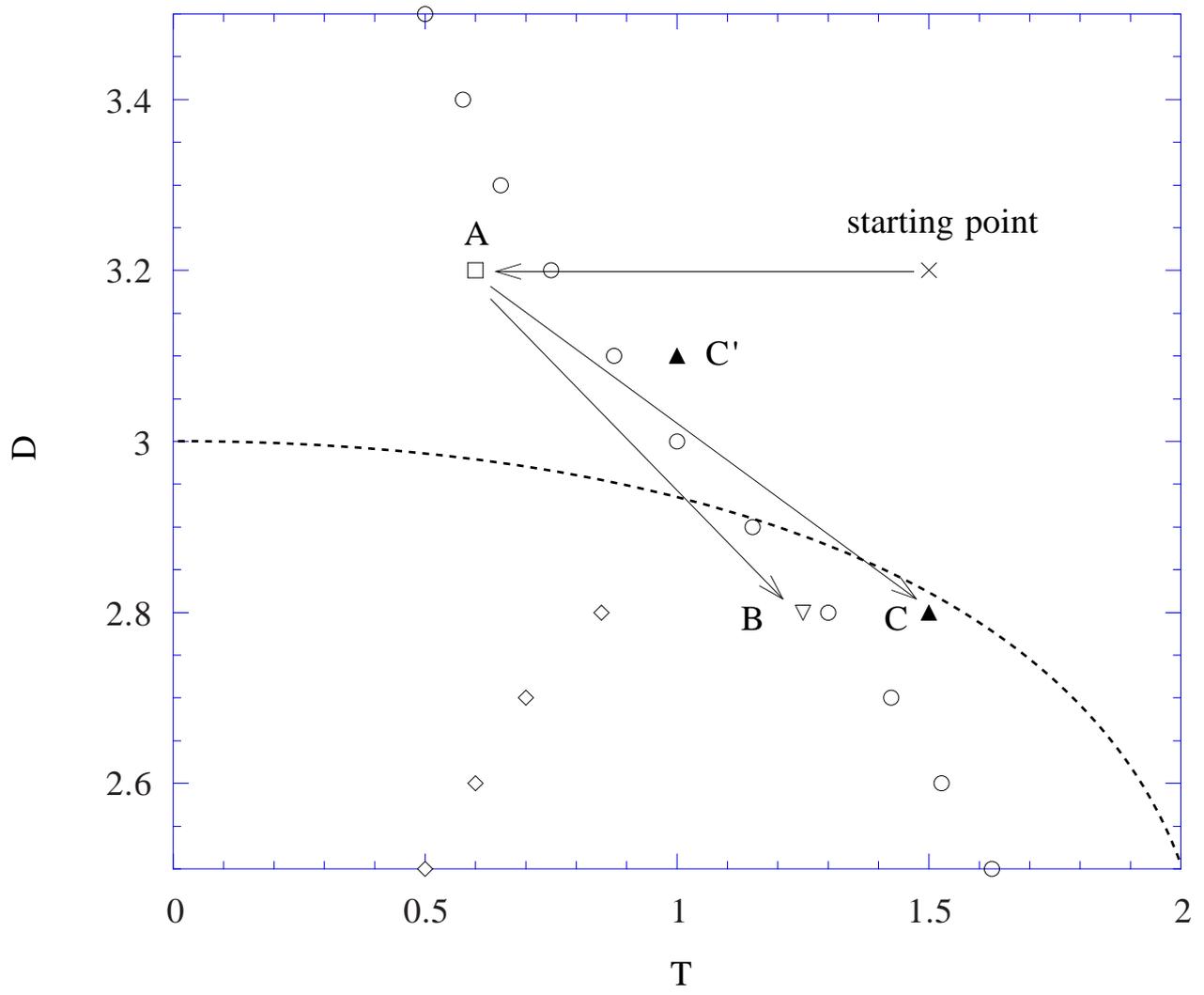
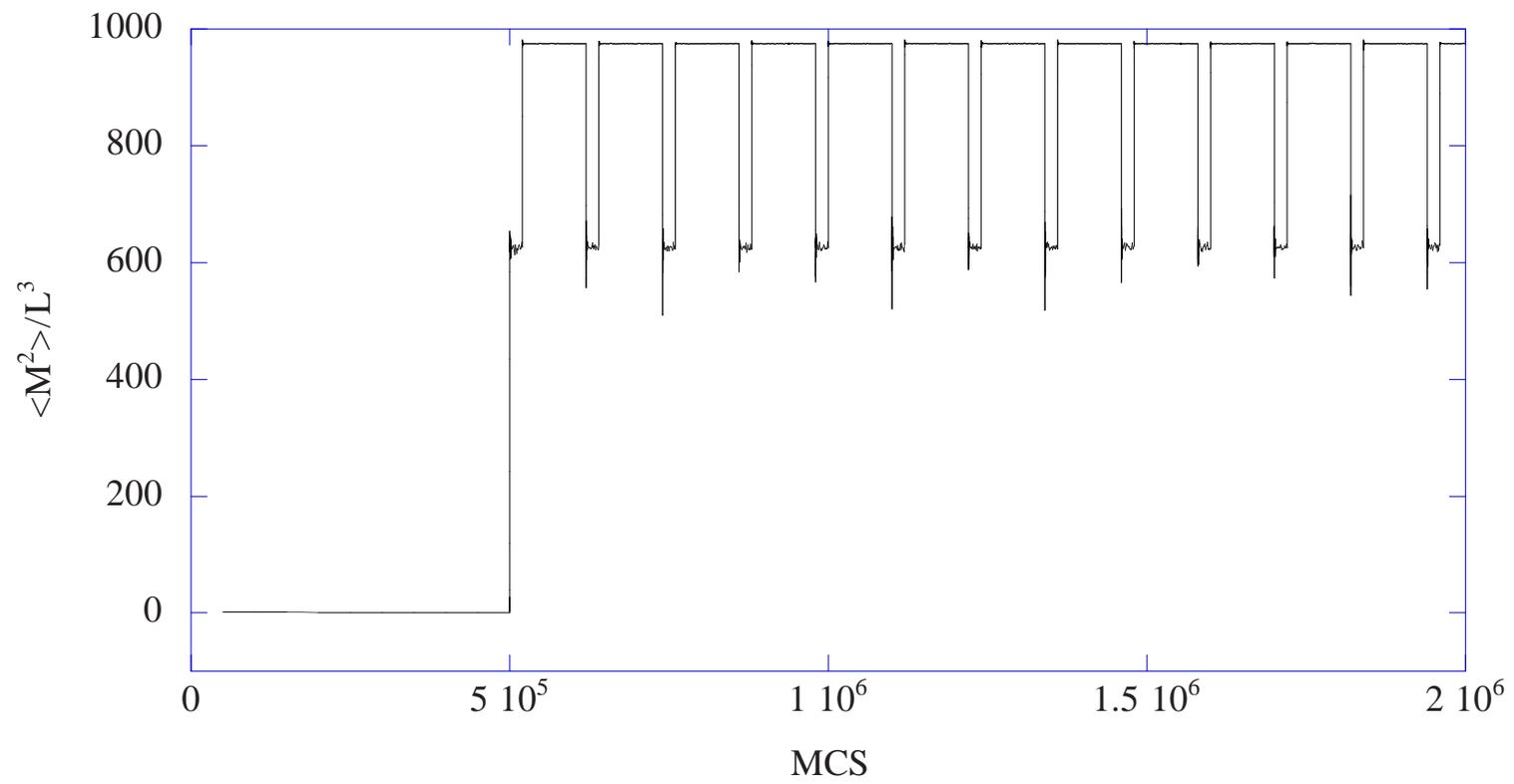
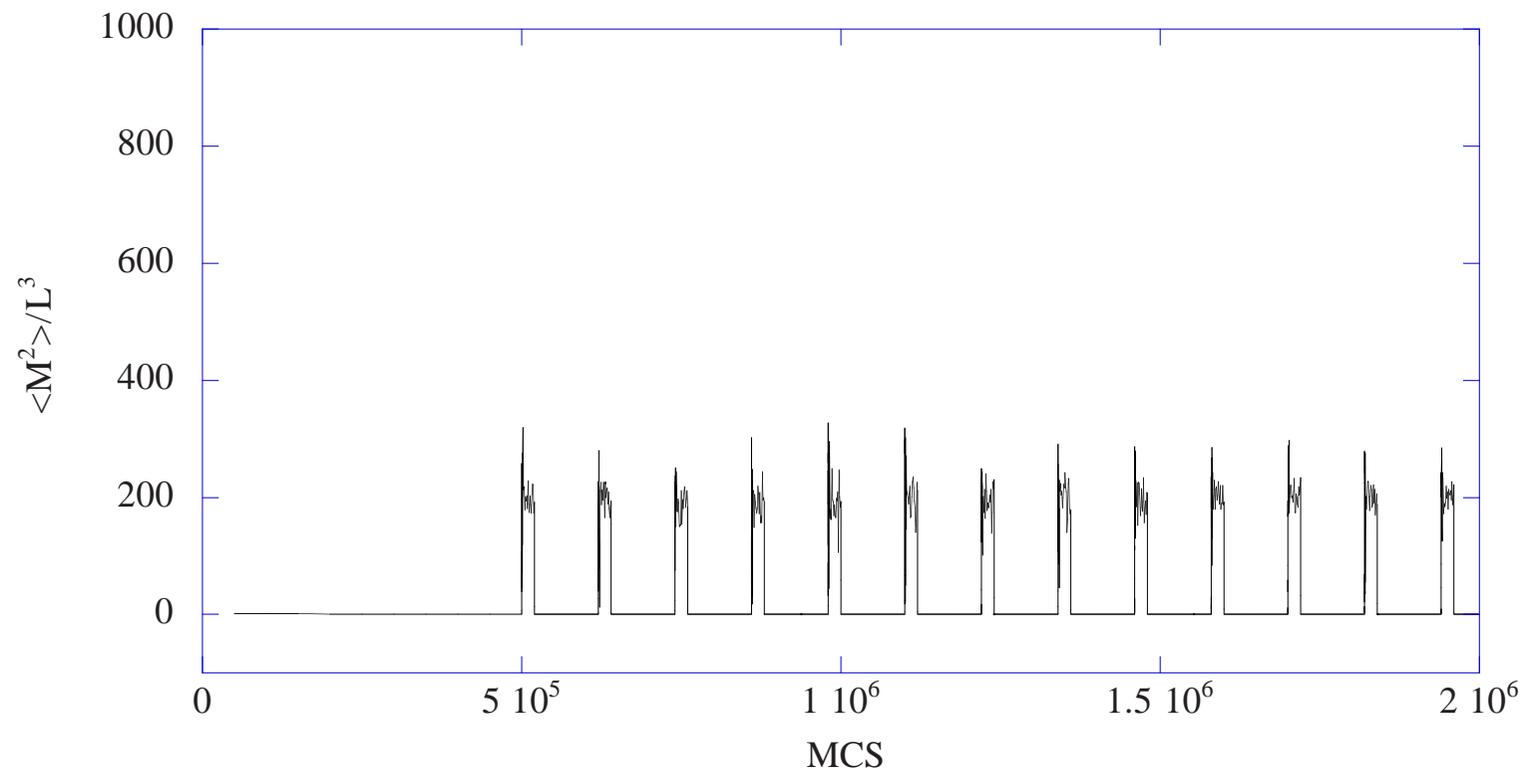
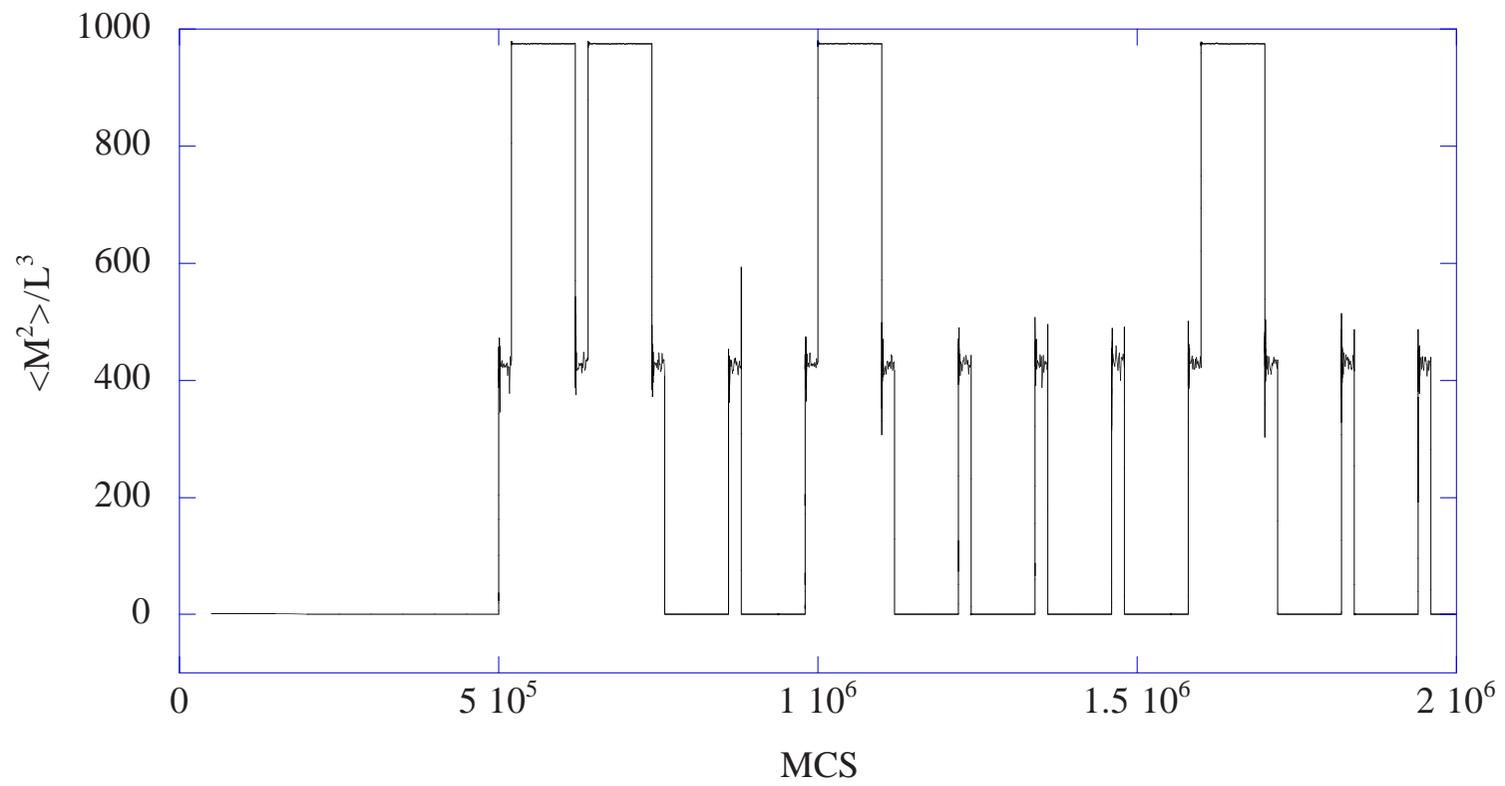
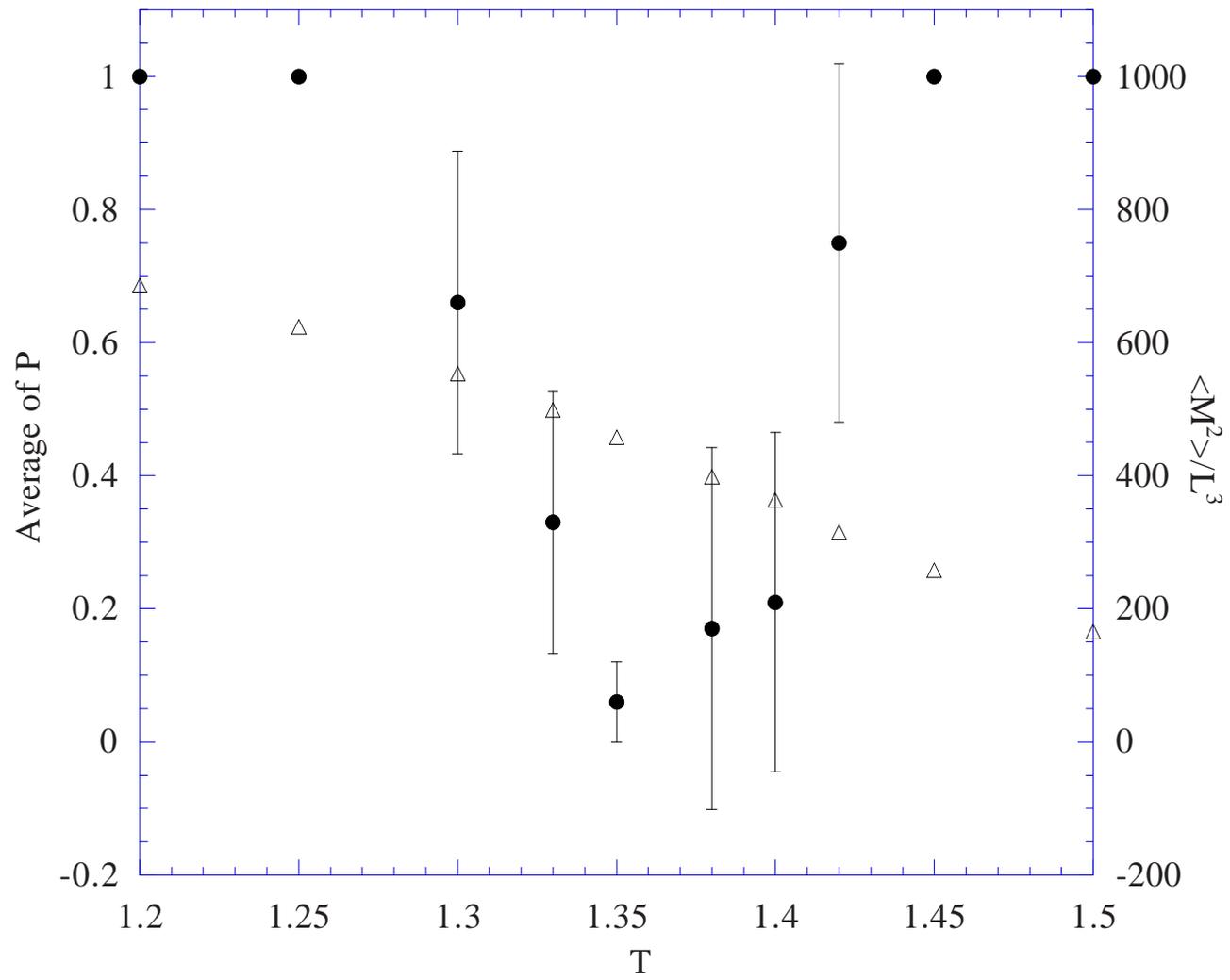
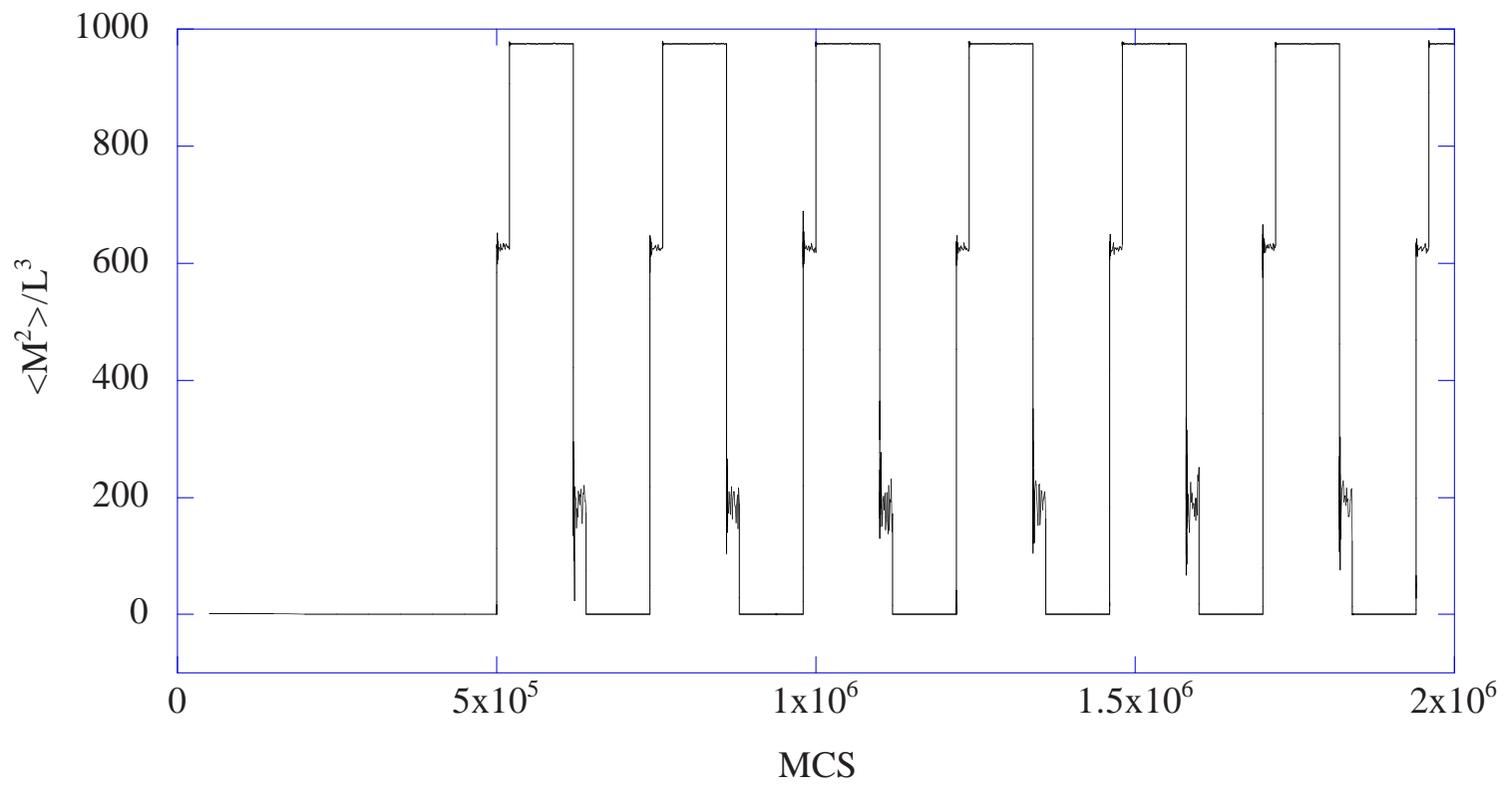
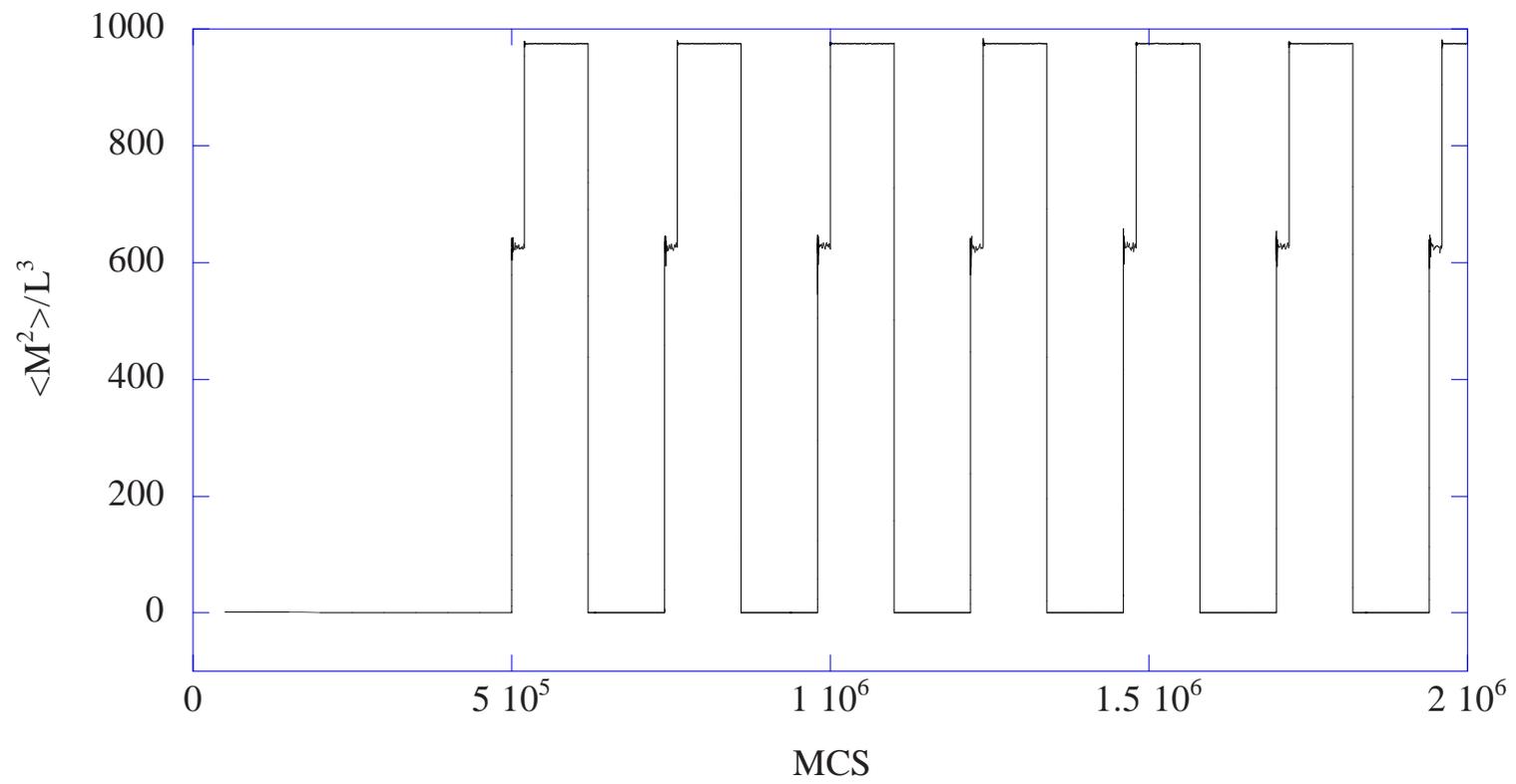
\end{document}